# On Kolmogorov Complexity of Random Very Long Braided Words

# V 2.

Dara O Shayda

dara@lossofgenerality.com

July 2013

## Abstract

Any positive word comprised of random sequence of tokens from a finite alphabet can be reduced (without change of length) using an appropriate size Braid group relationships. Surprisingly the Braid relationships dramatically reduce the Kolmogorov Complexity of the original random word and do so in distinct 'bands' of (rate of change) values with gaps in between. Distribution of these bands are estimated and empirical statistics collected by actually coding approximations to the Kolmogorov Complexity in *Mathematica* 9.0 . Lempel-Ziv-Welch lossless compression algorithm techniques used to estimate the distributions for the gaped bands. Evidence provided that such distributions of reduction in Kolmogorov Complexity based upon Braid groups are universal i.e. they can model more general algebraic structures other than Braid groups.

## Empirical Observation

At first glance a random <u>positive</u> long word, tokens from a finite alphabet, subjected to the substitutions of some appropriate Braid group is no further ordered nor disordered and therefore ought to stay as random as the original word. But actual symbolic experimentation on words of $10^6$ length shows a different view! Upon the substitutions of the Braid group relationships, even though the length stays the same, the approximated complexity of the reduced word drops drastically, and does so in gaps of values i.e. in discrete values and no occurrences in between.

At the beginning the author presumed that there was a computational inaccuracy some place in the code but as time went by a pattern of distinct drops in complexity of random VLBW (Very Long Braided Word) were repeatedly observed.

These observations coupled with symbolic calculations indicate a Braided structure found in very long random words. In other words the random signals in nature if interpreted as long Braided words unintuitively indicate discrete levels of information with specific distributions.

## Methodology of Investigation

Experimental code was used as rewrite system to reduce the words in different algebras. The results were checked against what is known in the literature therefore to make sure the code is sound.

While doing so Braided groups were tried and obtained uninteresting reduced words which looked as random as the original word.

Then more code was developed based upon Compress [ ] function in *Mathematica* to show that the Braids group relations do not alter the randomness of a long word by much. In doing so results were obtained to the contrary! Further code was developed to investigate the anomaly for long words of several thousands length.

The complexity measure approximations varied and then the author added more code to compute the Mean and Variance of the complexity measure. The investigation indicated that for Braid groups $\mathcal{B}_n$ of larger n, words must have at least length $10^6$ to obtain small Variance.

However original code consumed 8 GIG of physical memory and almost all the hard drive consumed by page-out memory, while reasonable CPU, on an IMAC to estimate complexity of $\mathcal{B}_{100}$ reductions for $10^6$ random long words. For a solution, new developed code made the memory consumption manageable by breaking the words into smaller words of length 20,000. For most words of length $10^6$ several hours (often 12+) of computation are necessary to get the proper output of reduced word. This is so due to the pattern matching algorithms necessary to find candidate sub-strings in the word for replacement.

Upon the results of the new code for such long words on larger n, suddenly a pattern of complexity reductions were noticeable in the Braid reduced random long words.

Since exact computation of Kolmogorov Complexity is non-existing, then a lossless less efficient computable function was used (Compress in *Mathematica*) but results were checked by the Mean and Variance of small collection of sample long random words.

The measures of such complexity were plotted and an unexpected familiar shape of $\frac{1}{n}$ was observed! Consequently a Fit function was used in *Mathematica* to find the best fit(s) to the said outputs.

From all above a theorem was formed, and again reusing the same methods parts of the proof were crafted which would have otherwise been impossible or taken much longer time or require much higher competency and teamwork to bring to fruition. (See [1,2])

# 1. Formalism

A free group $\mathcal{F}_n$ is a group with n finite generators $\{\sigma_1, \sigma_2, \cdots, \sigma_n\}$ and no relations between the generators. A word is a concatenation (product) of these generators e.g. $\sigma_2 \sigma_9^{-1} \sigma_{100}$. A positive word is a word with no negative powers i.e. inverse elements e.g. $\sigma_2^{120} \sigma_9^2 \sigma_{100}$.

A Braid group $\mathcal{B}_n$ is a group with n finite generators $\{\sigma_1, \sigma_2, \cdots, \sigma_n\}$ and the following two distinct lists of relations between the generators:

$$\sigma_i \sigma_{i+1} \sigma_i = \sigma_{i+1} \sigma_i \sigma_{i+1} \qquad\qquad 1 \leqslant i < n \qquad\qquad \text{(EQ 1.1)}$$
$$\sigma_i \sigma_j = \sigma_j \sigma_i \qquad\qquad\qquad\qquad |i - j| \geqslant 2 \qquad\qquad \text{(EQ 1.2)}$$

**Example 1.1 .** Braid group $\mathcal{B}_5$

$$\{\sigma_1, \sigma_2, \sigma_3, \sigma_4, \sigma_5\}$$

$$\{\sigma_1 \sigma_2 \sigma_1 \to \sigma_2 \sigma_1 \sigma_2, \ \sigma_2 \sigma_3 \sigma_2 \to \sigma_3 \sigma_2 \sigma_3, \ \sigma_3 \sigma_4 \sigma_3 \to \sigma_4 \sigma_3 \sigma_4, \ \sigma_4 \sigma_5 \sigma_4 \to \sigma_5 \sigma_4 \sigma_5,$$
$$\sigma_1 \sigma_3 \to \sigma_3 \sigma_1, \ \sigma_1 \sigma_4 \to \sigma_4 \sigma_1, \ \sigma_1 \sigma_5 \to \sigma_5 \sigma_1, \ \sigma_2 \sigma_4 \to \sigma_4 \sigma_2, \ \sigma_2 \sigma_5 \to \sigma_5 \sigma_2, \ \sigma_3 \sigma_5 \to \sigma_5 \sigma_3\}$$

Notation '$\to$' represents substitution i.e. left hand-side could be replaced by the right hand-side of the arrow and the value in the group is preserved. The author uses this notation vs. the '$=$' to indicate the computational 'process' of substitution.

**Remark 1.1**: *Use $\{\sigma_1, \sigma_2, \cdots\}$ for generators to match the textbooks and papers and as a unified set of tokens.*

By $w_G$ we mean the word formed by the generators of the group G and thus reducible according to its relations amongst the generators. At the same time we can again say $w_F$ which is the same word in yet another group with the same set of generators but different relations to reduce.

**Definition 1.1**: In specific case of the Braid groups $\mathcal{B}_n$, $w_{\mathcal{B}_n}$ should read as 'w is braided by $\mathcal{B}_n$' or '$w_{\mathcal{B}_n}$ is a braided word'. Vice-a-versa going from $w_{\mathcal{B}_n}$ to w should read as '$w_{\mathcal{B}_n}$ is unbraided'.

**Definition 1.2**: By $C\left(w_F\right)$ we mean the Kolmogorov Complexity of SOME fixed reduction(s) of w by the group relations of F.

The reductions could be many, so the context for what order of reductions has to be specified. Generally reductions are thought to be no further reducible, and even the latter might not be unique since the relations' order during the reduction could result in different irreducible values.

**Definition 1.3**: Quotient Kolmogorov Complexity with regards to group relations F and G is the following limit (if it exists)

$$\mathop{C}_{F,G} = \lim_{|w|\longrightarrow\infty} \frac{C\left(\frac{w}{F}\right) - C\left(\frac{w}{G}\right)}{C\left(\frac{w}{F}\right) - C\left(\underset{F}{const}\right)} \quad \text{where const is some constant word of size } |w|$$

Loosely speaking, this quotient defines a rate of change or a raw derivative for Kolmogorov Complexity based upon the variations in randomness caused by reduction from F to G.

**Remark 1.2**: *The subtraction* $C\left(\frac{w}{F}\right) - C\left(\underset{F}{const}\right)$ *is for getting rid of the constant* $c_{\mathcal{A}}$ *in Theorem A.1 .*

Sadly exact computation of $\mathop{C}_{F,G}$ is impossible, therefore a more statistical approach could possibly give good estimates:

**Definition 1.4**: Quotient Kolmogorov Complexity Estimator, given any fixed compression algorithm and the © function which first compresses the word accordingly and then outputs its length:

$$\mathop{C}_{F,G} \approx \left\langle \frac{©\left(\frac{w}{F}\right) - ©\left(\frac{w}{G}\right)}{©\left(\frac{w}{F}\right) - ©\left(\underset{F}{const}\right)} \right\rangle = \mathop{\hat{C}}_{F,G} \quad \text{for some set of very long words } w \in \{w_i\}$$

$\langle \rangle$ stands for Mean.

For all the above definitions and concepts we say $\underset{G}{w}$ is reduced 1-pass means the group relations were applied only once throughout the entire word w, m-pass means m times 'all' the relations were applied and m reductions accumulated, many-pass means a very large number of times the relations were applied.

Therefore we can say $\mathop{\hat{C}}_{F,G}$ is 1-pass estimator or m-pass or many-pass.

## 2. Statement And Discussion of Main Results (1-pass)

Our main objective is to compute bounds for:

$$\mathop{\hat{C}}_{F,G} = \left\langle \frac{©\left(\frac{w}{\mathcal{F}_n}\right) - ©\left(\frac{w}{\mathcal{G}_n}\right)}{©\left(\frac{w}{\mathcal{F}_n}\right) - ©\left(\underset{\mathcal{F}_n}{const}\right)} \right\rangle \quad n \geq 3 \qquad \text{(EQ 2.1)}$$

as a function of n for sets of very long random words w. Basically we take a random word over a finite alphabet i.e. a product in a free group $\mathcal{F}_n$ and then estimate its Kolmogorov Complexity $\copyright\left(\begin{matrix} w \\ \mathcal{F}_n \end{matrix}\right)$ . Then we Braid the word w and again estimate its Kolmogorov Complexity $\copyright\left(\begin{matrix} w \\ \mathcal{B}_n \end{matrix}\right)$ . Plug them into the (EQ 2.1) and obtain a rate of change for Kolmogorov Complexity going from a random word w to Braided w, repeat the process for different random w to compute the Mean.

Investigation started by coding $\copyright(w)$ which in *Mathematica* is

ByteCount[Compress[$w$]]      (EQ 2.2)

See [1].

Let's say n = 10, and we form a very long random word of $10^6$ length in group $\mathcal{F}_{10}$, all positive powers:

"$\sigma_7\sigma_1\sigma_7\sigma_9\sigma_6\sigma_7\sigma_3\sigma_6\sigma_4\sigma_2\sigma_9\sigma_3\sigma_5\sigma_9\sigma_9\sigma_1\sigma_8\sigma_2\sigma_9\sigma_8\sigma_4\sigma_7\sigma_8\sigma_6\sigma_{10}\sigma_3\sigma_3\sigma_9\sigma_7\sigma_5\sigma_2\sigma_1\sigma_8\sigma_6\sigma_9\sigma_2\sigma_5\sigma_9\sigma_5\cdots$"

Obviously there is no relations and therefore the word is a random sequence of tokens.

Then we form $\mathcal{B}_{10}$ the Braid group of 10 generators, and there are two sets of relations according to (EQ 1.1):

$\{\sigma_1\sigma_2\sigma_1 \to \sigma_2\sigma_1\sigma_2 ,\ \sigma_2\sigma_3\sigma_2 \to \sigma_3\sigma_2\sigma_3 ,\ \sigma_3\sigma_4\sigma_3 \to \sigma_4\sigma_3\sigma_4 ,\ \sigma_4\sigma_5\sigma_4 \to \sigma_5\sigma_4\sigma_5 ,$
$\sigma_5\sigma_6\sigma_5 \to \sigma_6\sigma_5\sigma_6 ,\ \sigma_6\sigma_7\sigma_6 \to \sigma_7\sigma_6\sigma_7 ,\ \sigma_7\sigma_8\sigma_7 \to \sigma_8\sigma_7\sigma_8 ,\ \sigma_8\sigma_9\sigma_8 \to \sigma_9\sigma_8\sigma_9 ,\ \sigma_9\sigma_{10}\sigma_9 \to \sigma_{10}\sigma_9\sigma_{10}\}$

and (EQ 1.2):

$\{\sigma_1\sigma_3 \to \sigma_3\sigma_1 ,\ \sigma_1\sigma_4 \to \sigma_4\sigma_1 ,\ \sigma_1\sigma_5 \to \sigma_5\sigma_1 ,\ \sigma_1\sigma_6 \to \sigma_6\sigma_1 ,\ \sigma_1\sigma_7 \to \sigma_7\sigma_1 ,\ \sigma_1\sigma_8 \to \sigma_8\sigma_1 ,$
$\sigma_1\sigma_9 \to \sigma_9\sigma_1 ,\ \sigma_1\sigma_{10} \to \sigma_{10}\sigma_1 ,\ \sigma_2\sigma_4 \to \sigma_4\sigma_2 ,\ \sigma_2\sigma_5 \to \sigma_5\sigma_2 ,\ \sigma_2\sigma_6 \to \sigma_6\sigma_2 ,\ \sigma_2\sigma_7 \to \sigma_7\sigma_2 ,$
$\sigma_2\sigma_8 \to \sigma_8\sigma_2 ,\ \sigma_2\sigma_{10} \to \sigma_9\sigma_2 ,\ \sigma_2\sigma_{10} \to \sigma_{10}\sigma_2 ,\ \sigma_3\sigma_5 \to \sigma_5\sigma_3 ,\ \sigma_3\sigma_6 \to \sigma_6\sigma_3 ,\ \sigma_3\sigma_7 \to \sigma_7\sigma_3 ,$
$\sigma_3\sigma_8 \to \sigma_8\sigma_3 ,\ \sigma_3\sigma_9 \to \sigma_9\sigma_3 ,\ \sigma_3\sigma_{10} \to \sigma_{10}\sigma_3 ,\ \sigma_4\sigma_6 \to \sigma_6\sigma_4 ,\ \sigma_4\sigma_7 \to \sigma_7\sigma_4 ,\ \sigma_4\sigma_8 \to \sigma_8\sigma_4 ,$
$\sigma_4\sigma_9 \to \sigma_9\sigma_4 ,\ \sigma_4\sigma_{10} \to \sigma_{10}\sigma_4 ,\ \sigma_5\sigma_7 \to \sigma_7\sigma_5 ,\ \sigma_5\sigma_8 \to \sigma_8\sigma_5 ,\ \sigma_5\sigma_9 \to \sigma_9\sigma_5 ,\ \sigma_5\sigma_{10} \to \sigma_{10}\sigma_5 ,$
$\sigma_6\sigma_8 \to \sigma_8\sigma_6 ,\ \sigma_6\sigma_9 \to \sigma_9\sigma_6 ,\ \sigma_6\sigma_{10} \to \sigma_{10}\sigma_6 ,\ \sigma_7\sigma_9 \to \sigma_9\sigma_7 ,\ \sigma_7\sigma_{10} \to \sigma_{10}\sigma_7 ,\ \sigma_8\sigma_{10} \to \sigma_{10}\sigma_8\}$

See [2].

Note that there are order to both sets and change of the order would change the outcome of substitutions. As it turns out the estimate $C_{\mathcal{F}_{10},\mathcal{B}_{10}}$ does change if we change the order of the relations listed above, but the changes are minute (See [1]).

**Remark 2.1**: *However it is best to randomize the order of the relations, in order to get a less biased more general view, See kBraidsList[ ] in [1].*

10 random words are generated, each with $10^6$ length and the following statistics collected:

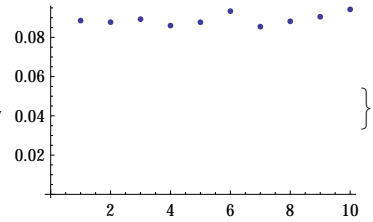

$\{0.0942366, 0.085456, 0.0890859, 0.00287523,$ $\}$

First number is Max, second number Min, third number Mean, and fourth number Standard Deviation. Therefore

$$\hat{C}_{\mathcal{F}_{10}, \mathcal{B}_{10}} = \left\langle \frac{\copyright\left(w_{\mathcal{F}_{10}}\right) - \copyright\left(w_{\mathcal{B}_{10}}\right)}{\copyright\left(w_{\mathcal{F}_{10}}\right) - \copyright\left(\text{const}_{\mathcal{F}_{10}}\right)} \right\rangle = 0.0890859 \text{ with error } \pm 0.00287523, \text{ in interval } [0.085456, \ 0.0942366]$$

kBraidsStat[ ] is the function that computes the above statistics, see [1].

Let's repeat the experiment for n = 11, bumped by 1:

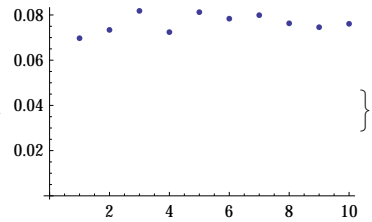

$\{0.0818024, 0.0697132, 0.0763784, 0.00397013,$ $\}$

It is obvious that the two values are gaped and separated i.e. discrete in nature:

$$\hat{C}_{\mathcal{F}_{11}, \mathcal{B}_{11}} = \left\langle \frac{\copyright\left(w_{\mathcal{F}_{11}}\right) - \copyright\left(w_{\mathcal{B}_{11}}\right)}{\copyright\left(w_{\mathcal{F}_{11}}\right) - \copyright\left(\text{const}_{\mathcal{F}_{11}}\right)} \right\rangle = 0.0763784 \text{ with error } \pm 0.00397013, \text{ in interval } [0.0697132, \ 0.0818024]$$

See [1] for more statistics.

Collecting a few such statistics, for words with $10^6$ length, and n between [3, 100] a definite shape for the distribution of $C_{\mathcal{F}_n, \mathcal{B}_n}$ is uncovered with the parametric fit:

$$C_{\mathcal{F}_n, \mathcal{B}_n} = 0.00837797 + \frac{0.572318}{n} \quad n \in [3, 100]$$

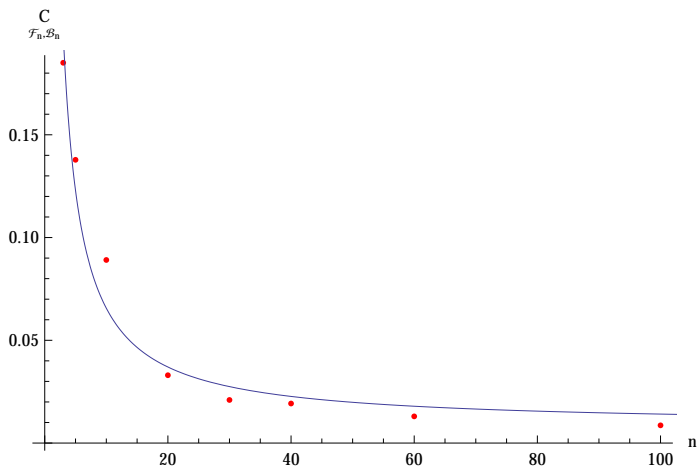

Note that the red dots are the collected statistics, the blue curve is the putative fit function (See [1]).

Now we are equipped to state the main result:

**VLBW Theorem 2.1**: For finite sets of very long random words $w \in \{w_i\}$, $\hat{C}_{\mathcal{F}_n, \mathcal{B}_n} = \left( \frac{\mathbb{C}\left(\frac{w}{\mathcal{F}_n}\right) - \mathbb{C}\left(\frac{w}{\mathcal{B}_n}\right)}{\mathbb{C}\left(\frac{w}{\mathcal{F}_n}\right) - \mathbb{C}\left(\text{const}\right)} \right)$ 1-pass estimator converges to monotonically decreasing gaped discrete values for $n \geqslant 3$, $\hat{C}_{\mathcal{F}_n, \mathcal{B}_n}$ depends only on 'n' and $0 < \hat{C}_{\mathcal{F}_n, \mathcal{B}_n} < 1$ assuming $\mathbb{C}$ based on a lossless flavor of LZW compression algorithm. Upper bound for distribution is $O\left(\frac{1}{n}\right)$.

LZW stands for Lempel-Ziv-Welch. See Appendix B for examples of LZW to understand the algorithm.

**Remark 2.2**: *Obviously any flavor of LZW is far of the goal for approximating Kolmogorov Complexity, but the* $\hat{C}_{\mathcal{F}_n, \mathcal{B}_n}$ *estimates are good enough to describe the same situations as if exact Kolmogorov Complexity was used.*

## 3. Proof

**Sketch**:

1. Allocate a buffer containing a list of non-random repeated strings in w, we do not need to worry about how this list was uncovered or stored e.g. 'read from left to write' or 'prefix words' and other such considerations. All these at worst will be covered by an additional term $O(1)$ which is canceled out or divided by a large nume at the end.

2. To each entry in the buffer assign a code or token formed from an alternative alphabet. Alternative alphabet is a suggestion to make the below estimates easier to understood, but it is not essential.

3. Replace the repeated sub-strings of w with codes/tokens from #2, form a new word (to output).

4. <u>Concatenate the #1 and #2 and this would be a candidate (approximation) for the shortest possible program (See Appendix A) which could output (reconstruct by lossless decompression) w in a lossless fashion.</u>

5. Provide constraints to assure the length of #4 to be shorter than length of w

6. Show $\dfrac{\mathbb{C}\left(w_{\mathcal{F}_n}\right) - \mathbb{C}\left(w_{\mathcal{B}_n}\right)}{\mathbb{C}\left(w_{\mathcal{F}_n}\right) - \mathbb{C}\left(\underset{\mathcal{F}_n}{\text{const}}\right)}$ is independent of w as length of w increases and only depends on n and $n \ll |w|$

7. Estimate #6 as quotient polynomials and logs

**Remark 3.1**: *Steps 1-4 are raw parts of LZW compression without the worries about the prefixed words and so on, since we are seeking estimates for very long w plugged into* $\dfrac{\mathbb{C}\left(w_{\mathcal{F}_n}\right) - \mathbb{C}\left(w_{\mathcal{B}_n}\right)}{\mathbb{C}\left(w_{\mathcal{F}_n}\right) - \mathbb{C}\left(\underset{\mathcal{F}_n}{\text{const}}\right)}$. *See Appendix B for a short example by Welch to render steps 1-3.*

For an untrained eye it is rather difficult to see any repeated patterns in $w_{\mathcal{B}_n}$ after reductions. For the sake of clarity let's work with $w_{\mathcal{B}_{10}}$ for n = 10, to develop an insight into the matter of repeated strings of $w_{\mathcal{B}_n}$.

Sample word w, basically a random word in a free group $\mathcal{F}_{10}$:

"$\sigma_7\sigma_1\sigma_7\sigma_9\sigma_6\sigma_7\sigma_3\sigma_6\sigma_4\sigma_2\sigma_9\sigma_3\sigma_5\sigma_9\sigma_9\sigma_1\sigma_8\sigma_2\sigma_9\sigma_8\sigma_4\sigma_7\sigma_8\sigma_6\sigma_{10}\sigma_3\sigma_3\sigma_9\sigma_7\sigma_5\sigma_2\sigma_1\sigma_8\sigma_6\sigma_9\sigma_2\sigma_5\sigma_9\sigma_5\cdots$"

With very short repeated patterns:

$\sigma_{10}\ \sigma_9\ \sigma_4\ \sigma_1$

$\sigma_9\ \sigma_6\ \sigma_3$

Strings with decreasing index are at most length 4.

Switch to $\mathcal{B}_{10}$ which has two lists of relations for reduction of the words in its group structure (See [1,2]).

List 1

$\{\sigma_1\sigma_2\sigma_1 \rightarrow \sigma_2\sigma_1\sigma_2\,,\ \sigma_2\sigma_3\sigma_2 \rightarrow \sigma_3\sigma_2\sigma_3\,,\ \sigma_3\sigma_4\sigma_3 \rightarrow \sigma_4\sigma_3\sigma_4\,,\ \sigma_4\sigma_5\sigma_4 \rightarrow \sigma_5\sigma_4\sigma_5\,,$
$\ \sigma_5\sigma_6\sigma_5 \rightarrow \sigma_6\sigma_5\sigma_6\,,\ \sigma_6\sigma_7\sigma_6 \rightarrow \sigma_7\sigma_6\sigma_7\,,\ \sigma_7\sigma_8\sigma_7 \rightarrow \sigma_8\sigma_7\sigma_8\,,\ \sigma_8\sigma_9\sigma_8 \rightarrow \sigma_9\sigma_8\sigma_9\,,\ \sigma_9\sigma_{10}\sigma_9 \rightarrow \sigma_{10}\sigma_9\sigma_{10}\}$

Sample non-random repeated strings generated by List 1 set of relations, the indices are decreasing:

$\sigma_{10}\ \sigma_{10}\ \sigma_4\ \sigma_3\ \sigma_1\ \sigma_1$

$\sigma_{10}\ \sigma_4\ \sigma_4\ \sigma_4\ \sigma_1$

$\sigma_9\ \sigma_9\ \sigma_2\ \sigma_1\ \sigma_3$

$\sigma_{10}\ \sigma_{10}\ \sigma_2$

Therefore the application of the relations in List 1 cause longer decreasing indices for repeated strings i.e. reduction in randomness.

List 2

$\{\sigma_1\sigma_3 \rightarrow \sigma_3\sigma_1,\ \sigma_1\sigma_4 \rightarrow \sigma_4\sigma_1,\ \sigma_1\sigma_5 \rightarrow \sigma_5\sigma_1,\ \sigma_1\sigma_6 \rightarrow \sigma_6\sigma_1,\ \sigma_1\sigma_7 \rightarrow \sigma_7\sigma_1,\ \sigma_1\sigma_8 \rightarrow \sigma_8\sigma_1,$
$\sigma_1\sigma_9 \rightarrow \sigma_9\sigma_1,\ \sigma_1\sigma_{10} \rightarrow \sigma_{10}\sigma_1,\ \sigma_2\sigma_4 \rightarrow \sigma_4\sigma_2,\ \sigma_2\sigma_5 \rightarrow \sigma_5\sigma_2,\ \sigma_2\sigma_6 \rightarrow \sigma_6\sigma_2,\ \sigma_2\sigma_7 \rightarrow \sigma_7\sigma_2,$
$\sigma_2\sigma_8 \rightarrow \sigma_8\sigma_2,\ \sigma_2\sigma_9 \rightarrow \sigma_9\sigma_2,\ \sigma_2\sigma_{10} \rightarrow \sigma_{10}\sigma_2,\ \sigma_3\sigma_5 \rightarrow \sigma_5\sigma_3,\ \sigma_3\sigma_6 \rightarrow \sigma_6\sigma_3,\ \sigma_3\sigma_7 \rightarrow \sigma_7\sigma_3,$
$\sigma_3\sigma_8 \rightarrow \sigma_8\sigma_3,\ \sigma_3\sigma_9 \rightarrow \sigma_9\sigma_3,\ \sigma_3\sigma_{10} \rightarrow \sigma_{10}\sigma_3,\ \sigma_4\sigma_6 \rightarrow \sigma_6\sigma_4,\ \sigma_4\sigma_7 \rightarrow \sigma_7\sigma_4,\ \sigma_4\sigma_8 \rightarrow \sigma_8\sigma_4,$
$\sigma_4\sigma_9 \rightarrow \sigma_9\sigma_4,\ \sigma_4\sigma_{10} \rightarrow \sigma_{10}\sigma_4,\ \sigma_5\sigma_7 \rightarrow \sigma_7\sigma_5,\ \sigma_5\sigma_8 \rightarrow \sigma_8\sigma_5,\ \sigma_5\sigma_9 \rightarrow \sigma_9\sigma_5,\ \sigma_5\sigma_{10} \rightarrow \sigma_{10}\sigma_5,$
$\sigma_6\sigma_8 \rightarrow \sigma_8\sigma_6,\ \sigma_6\sigma_9 \rightarrow \sigma_9\sigma_6,\ \sigma_6\sigma_{10} \rightarrow \sigma_{10}\sigma_6,\ \sigma_7\sigma_9 \rightarrow \sigma_9\sigma_7,\ \sigma_7\sigma_{10} \rightarrow \sigma_{10}\sigma_7,\ \sigma_8\sigma_{10} \rightarrow \sigma_{10}\sigma_8\}$

By reductions of List 2, more longer decreasing indices are found for repeated strings:

$\sigma_{10}\ \sigma_{10}\ \sigma_{10}\ \sigma_7\ \sigma_7\ \sigma_7\ \sigma_5\ \sigma_3\ \sigma_3\ \sigma_1$

$\sigma_{10}\ \sigma_{10}\ \sigma_{10}\ \sigma_8\ \sigma_6\ \sigma_6\ \sigma_6\ \sigma_4\ \sigma_1$

$\sigma_{10}\ \sigma_{10}\ \sigma_8\ \sigma_7\ \sigma_7\ \sigma_5\ \sigma_4\ \sigma_3$

$\sigma_{10}\ \sigma_{10}\ \sigma_8\ \sigma_5\ \sigma_5\ \sigma_2\ \sigma_2$

$\sigma_{10}\ \sigma_{10}\ \sigma_{10}\ \sigma_5\ \sigma_5\ \sigma_2$

$\sigma_6\ \sigma_5\ \sigma_1\ \sigma_1\ \sigma_1$

$\sigma_8\ \sigma_1\ \sigma_1\ \sigma_1$

$\sigma_7\ \sigma_6\ \sigma_6$

In conclusion we observed that Braided relations induce new and longer non-random repeated patterns in long random words! Thus reduction in Kolmogorov Complexity is possible though we need to establish that this reduction is substantial. It could have been that there are such reductions in randomness but would not reflect much change in Kolmogorov Complexity.

**Step 1**:

Let's allocate a buffer for these non-random repeated strings:

**a**) Length of the buffer set to be buffLen = $h(n)$ where $h(n)$ is a polynomial in powers of n and log $n$
**b**) Width of the buffer i.e. the length of the non-random repeated strings set to be
buffWidth = $\log_2(n)\ p(n)$ where $p(n)$ is a polynomial in powers of n and log $n$

Obviously the following inequality has to hold in order for compression to be substantial enough:

$$\text{buffLen} \times \text{buffWidth} = \log_2(n)\, p(n)\, h(n) \ll \log_2(n)\, |w| \quad \text{(EQ 3.1)}$$

**Step 2**:

**c**) To each entry in the said buffer assign a code or a token from an alphabet e.g. $\Psi_i$ $i = 1, 2 \cdots h(n)$. Therefore the repeated strings are entirely replaced by one token each and that is how the compression outputs the compressed form.

**d**) Some of these are the single letter alphabets of original w with the constraint:

$$\text{buffLen} = h(n) > n \quad \text{(EQ 3.2)}$$

This in general is not true, imagine concatenating a word to itself, with buffer of length 1 i.e. the word itself algorithm can achieve 50% compression. But this cannot happen in our specific case since the word is assumed to be random.

**Step 3**:

**e**) Replace the new tokens from Step 2 into the w and therefore we get a shorter word of length:

$$\log_2(h(n))\, \frac{|w|}{p(n)} \quad \text{(EQ 3.3)}$$

Basically $\log_2(h(n))\, \frac{1}{p(n)} < 1$ which is desired to compress the length of the output.

**Step 4-5**:

**f**) Concatenate the buffer and the new shorter word (this would be the complete output word)
**g**) Compute the total length of the said concatenation and it has to be considerably less than length of original w:

$$\log_2(n)\, p(n)\, h(n) + \log_2(h(n))\, \frac{|w|}{p(n)} < \log_2(n)\, |w| \quad \text{(EQ 3.4)}$$

Eliminate |w| and assume it being very large :

$$0 < \log_2(n) - \frac{\log_2(h(n))}{p(n)} \quad \text{or} \quad 0 < \frac{\log_2(h(n))}{\log_2(n)\, p(n)} < 1 \quad \text{(EQ 3.5)}$$

**Step 6**:

**h**) Compute estimations for $\dfrac{\copyright\left(\frac{w}{\mathcal{F}_n}\right) - \copyright\left(\frac{w}{\mathcal{B}_n}\right)}{\copyright\left(\frac{w}{\mathcal{F}_n}\right) - \copyright\left(\frac{\text{const}}{\mathcal{F}_n}\right)}$ term by term:

$$\copyright\left(\frac{w}{\mathcal{F}_n}\right) \approx \log_2(n)\, |w| \quad \text{(EQ 3.6)} \quad \text{since w is assumed to be random}$$

$$\copyright\left(\frac{w}{\mathcal{B}_n}\right) \approx \log_2(n)\, p(n)\, h(n) + \log_2(h(n))\, \frac{|w|}{p(n)} \quad \text{(EQ 3.7)}$$

$$\underset{\mathcal{F}_n}{\text{©}}\left(\text{const}\right) \approx \log_2(n)\,\log_2(|w|) \quad \text{(EQ 3.8)}$$

Therefore

$$\frac{\underset{\mathcal{F}_n}{\text{©}}\left(w\right) - \underset{\mathcal{B}_n}{\text{©}}\left(w\right)}{\underset{\mathcal{F}_n}{\text{©}}\left(w\right) - \underset{\mathcal{F}_n}{\text{©}}\left(\text{const}\right)} \approx \frac{\log_2(n)\,|w| - \log_2(n)\,p(n)\,h(n) - \log_2(h(n))\,\frac{|w|}{p(n)}}{\log_2(n)\,|w| - \log_2(n)\,\log_2(|w|)} \approx 1 - \frac{\log_2(h(n))}{\log_2(n)\,p(n)} < 1 \quad \text{(EQ 3.9)}$$

Note that $\frac{\log_2(|w|)}{|w|} \longrightarrow 0$ as $|w|$ gets larger since it is assumed w is a long word. Right hand side of (EQ 3.9) is true since (EQ 3.5) is true.

(EQ 3.9) proves that $\underset{\mathcal{F}_n,\mathcal{B}_n}{\hat{C}}$ varies only by parameter n and $0 < \underset{\mathcal{F}_n,\mathcal{B}_n}{\hat{C}} < 1$ .

**Step 7**:

Start by

$$\log_2(n)\,p(n) = \log_2\left(n^{p(n)}\right) \quad \text{(EQ 3.10)}$$

And (EQ 3.5)

$$\frac{\log_2(h(n))}{\log_2(n)\,p(n)} < 1 \implies \frac{\log_2(h(n))}{\log_2(n^{p(n)})} < 1 \implies \log_2(h(n)) < \log_2\left(n^{p(n)}\right) \implies h(n) < n^{p(n)} \quad \text{(EQ 3.11)}$$

And by taking into account (EQ 3.2)

$$h(n) \in \left[n,\ n^{p(n)}\right) \quad \text{(EQ 3.12)}$$

From (EQ 3.9, 3.5, 3.11) we need to minimize the middle term:

$$0 < \frac{\log_2(h(n))}{\log_2(n)\,p(n)} < 1$$

In order to find the shortest program of the length assuming $n \ll |w|$ :

$$\log_2(n)\,p(n)\,h(n) + \log_2(h(n))\,\frac{|w|}{p(n)}$$

<u>What is chosen for p(n) and h(n) have to be based upon the relations of the particular group in question i.e. $\mathcal{B}_n$ to minimize the shortest program.</u>

Then let's calculate the distribution SPECIFIC to Braid groups.

We know that from Braid groups monotonically decreasing sequences are formed as seen earlier above (STEP 1), and they are sequences maxed at length c with possible repetitions of single tokens, see [6]:

$p(n) = c = O(1)$     (ignoring some constant that might be needed for repetitions) (EQ 3.13)

And set h(n) to the largest possible value i.e. largest possible number of repetitious words, while maintaining the inequality (EQ 3.11) or $h(n) < n^c$ :

$h(n) = n^{c(1-\delta)}$   $0 < \delta < 1$     (EQ 3.14)

Then we get the distribution up to order of magnitude:

$1 - \frac{\log_2(h(n))}{\log_2(n)\,p(n)} = 1 - \frac{\log_2(n^{c(1-\delta)})}{\log_2(n)\,c} = 1 - \frac{\log_2(n)\,c\,(1-\delta)}{\log_2(n)\,c} = \delta$     (EQ 3.15)

Assuming $\delta$ is smallest such order of magnitude (remember $1 - \frac{1}{\log_2(n)} < 1 - \frac{1}{n}$ therefore $O\left(\frac{1}{\log_2(n)}\right)$ will not do) :

$\delta < \frac{c'}{n} = O\left(\frac{1}{n}\right)$  for some constant c'   (EQ 3.16)

And as n increases  $\hat{C}_{\mathcal{F}_n, \mathcal{B}_n} \longrightarrow 0$ .

$\square$

Now for a GENERAL order of magnitude for other than Braid groups:

1. Set the p(n) to any reasonable integer function
2. Replace (EQ 3.13) with p(n)
3. Keep h(n) almost-maxed as $O(n^{p(n)(1-\delta)})$ in (EQ 3.14)
4. Set $\delta = \frac{1}{q(n)}$

And we immediately get the following Corollary:

**Corollary 3.1**: For any G with group relations reduction endowed with the property

$\left|\frac{W}{G}\right| = \left|\frac{W}{\mathcal{F}_n}\right|$  for free group $\mathcal{F}_n$

then

$\hat{C}_{\mathcal{F}_n, G} = O\left(\frac{1}{q(n)}\right) = \hat{C}_{\mathcal{F}_{q(n)}, \mathcal{B}_{q(n)}}$

assuming w being a positive random long word and some appropriate monotonically increasing integer function q(n).

In other words $\hat{C}_{\mathcal{F}_n, \mathcal{B}_n}$ provides a <u>universal distribution</u> for estimations of Quotient Kolmogorov Complexity.

## 4. Many-Pass (under investigation)

Further code experiments indicated that the 100-pass reductions produced a Gamma or Poisson looking distribution, see [7]:

$$0.08 + 0.045438 \, e^{-0.128756 \, n} \, n^{1.48723}$$

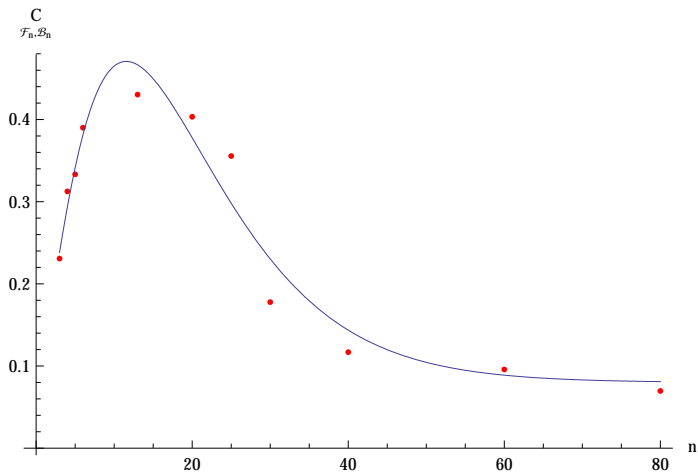

The right-most graph shows the distribution of the decreasing strings (decreasing in indices), 500 such strings, randomly chosen at the middle of the long word which is 30,000 in length in group $\mathcal{F}_{15}$. These strings start with the largest index number i.e. 15 and end when the decreasing sequence halts with a non-decreasing index. The plots are less than 500 since the length 1 strings were not accounted for. Index repetition allowed.

Example:

Word:

$$\sigma_2 \, \sigma_3 \, \sigma_1 \, \sigma_1 \, \sigma_2 \, \sigma_3 \, \sigma_3 \, \sigma_1 \, \sigma_2 \, \sigma_3$$

Decreasing string with repetition

$$\sigma_3 \, \sigma_1 \, \sigma_1 \quad \text{or} \quad \sigma_3 \, \sigma_3 \, \sigma_1 \, \sigma_2$$

Note that the second index 2 is a decreasing string of length 1.

Middle graph is frequency of occurrence of the said string in the long word.

The series of number, See [6]:

144 Max for repeated frequency of occurrence by the 500 strings
2 Min for the repeated  frequency of occurrence
71.7351  Mean for the  repeated frequency of occurrence
61.4 Standard deviation for the repeated frequency of occurrence

$\{$144., 2., 71.7351, 61.4, 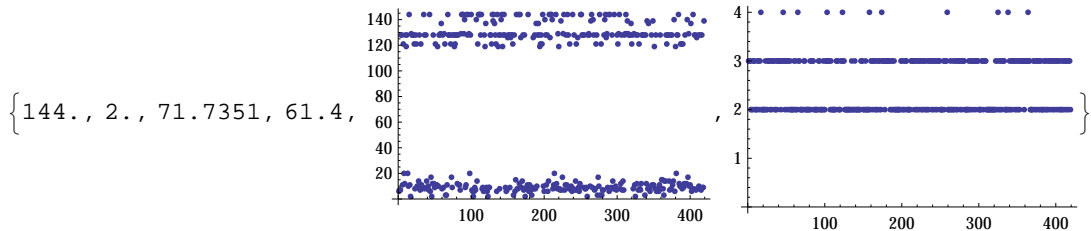 $\}$

Then the $\mathcal{B}_{15}$ reductions were computed, 1-pass only:

$\{$257., 2., 45.5124, 74.8104, 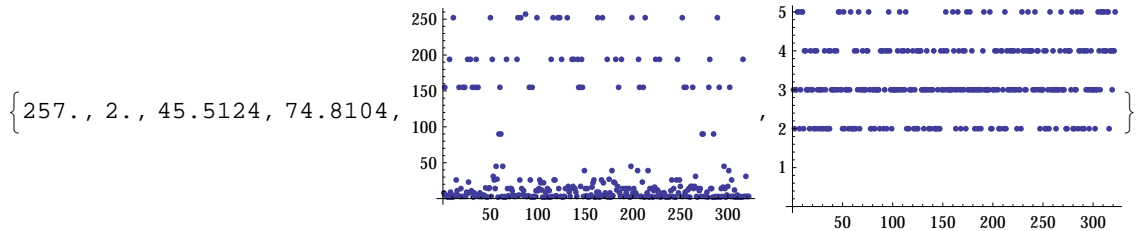 $\}$

Then the $\mathcal{B}_{15}$ reductions were computed, 1000000-pass:

$\{$260., 2., 60.8711, 74.8805, 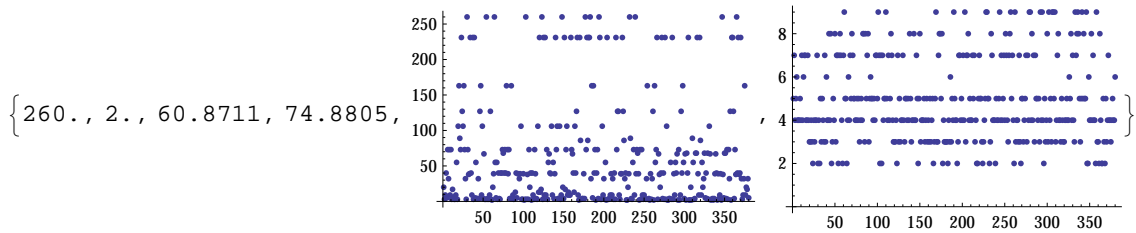 $\}$

# Appendix A

We assume all strings and programs are binary coded.

**Definition A.1**: The Kolmogorov Complexity $C_\mathcal{U}(x)$ of a string x with respect to a universal computer (Turing Machine) $\mathcal{U}$ is defined as

$$C_\mathcal{U}(x) = \min_{p:\mathcal{U}(p)=x} l(p)$$

the minimum length program p in $\mathcal{U}$ which outputs x.

**Theorem A.1 (Universality of the Kolmogorov Complexity)**: *If $\mathcal{U}$ is a universal computer, then for any other computer $\mathcal{A}$ and all strings x,*

$$C_\mathcal{U}(x) \le C_\mathcal{A}(x) + c_\mathcal{A}$$

where the constant $c_\mathcal{A}$ does not depend on x.

**Corollary A.1**: $\lim\limits_{l(x)\to\infty} \frac{C_\mathcal{U}(x) - C_\mathcal{A}(x)}{l(x)} = 0$ *for any two universal computers.*

**Remark A.1**: *Therefore we drop the universal computer subscript and simply write C(x).*

**Theorem A.2**: *$C(x) \le |x| + c$.*

A string x is called incompressible if $C(x) \ge x$ .

**Definition A.2**: Self-delimiting string (or program) is a string or program which has its own length encoded as a part of itself i.e. a Turing machine reading Self-delimiting string knows exactly when to stop reading.

**Definition A.3**: The Conditional or Prefix Kolmogorov Complexity of self-delimiting string x given string y is

$$K(x \mid y) = \min_{p:\mathcal{U}(p,y)=x} l(p)$$

The length of the shortest program that can compute both x and y and a way to tell them apart is

$$K(x, y) = \min_{p:\mathcal{U}(p)=x,y} l(p)$$

**Remark A.2**: *x, y can be thought of as concatenation of the strings with additional separation information.*

**Theorem A.3**: $K(x) \le l(x) + 2 \log l(x) + O(1), \quad K(x \mid l(x)) \le l(x) + O(1)$ .

**Theorem A.4**: $K(x, y) \le K(x) + K(y)$ .

**Theorem A.5**: $K(f(x)) \leq K(x) + K(f)$ , $f$ *a* computble function

# Appendix B

See [3]

**Table 3.** A string table for the example in Figure 5. The string table is initialized with three code values for the three characters, shown above the dotted line. Code values are assigned in sequence to new strings.

| STRING TABLE | | ALTERNATE TABLE | |
|---|---|---|---|
| a | 1 | a | 1 |
| b | 2 | b | 2 |
| c | 3 | c | 3 |
| ab | 4 | 1b | 4 |
| ba | 5 | 2a | 5 |
| abc | 6 | 4c | 6 |
| cb | 7 | 3b | 7 |
| bab | 8 | 5b | 8 |
| baba | 9 | 8a | 9 |
| aa | 10 | 1a | 10 |
| aa | 11 | 10a | 11 |
| aaaa | 12 | 11a | 12 |

| INPUT SYMBOLS | a | b | a | b | c | b | a | b | a | b | a | a | a | a | a | a |
|---|---|---|---|---|---|---|---|---|---|---|---|---|---|---|---|---|
| OUTPUT CODES | 1 | 2 | | 4 | | 3 | 5 | | | 8 | 1 | 10 | | | 11 | |
| | | | 5 | | | | 7 | | | 9 | | | | 11 | | |
| NEW STRING ADDED TO TABLE | | 4 | | | 6 | | | 8 | | | 10 | | | 12 | | |

**Figure 5.** A compression example. The input data, being read from left to right, is examined starting with the first character a. Since no matching string longer than a exists in the table, the code 1 is output for this string and the extended string ab is put in the table under code 4. Then b is used to start the next string. Since its extension ba is not in the table, it is put there under code 5, the code for b is output, and a starts the next string. This process continues straightforwardly.

| INPUT CODES | 1 | 2 | 4 | 3 | 5 | 8 | 1 | 10 | 11 |
|---|---|---|---|---|---|---|---|---|---|
| | ∨ | ∨ | ∨ | ∨ | ∨ | ∨ | ∨ | ∨ | ∨ |
| | a | b | 1b | c | 2a | 5b | a | 1a | 10a |
| | | | ∨ | | ∨ | ∨ | | ∨ | ∨ |
| | | | a | | b | 2a | | a | 1a |
| | | | | | | ∨ | | | ∨ |
| | | | | | | b | | | a |
| | | | | | | | | | ∨ |
| OUTPUT DATA | a | b | ab | c | ba | bab | a | aa | aaa |
| STRING ADDED TO TABLE | | 4 | | | 6 | | 8 | | 10 |
| | | | 5 | | | 7 | | 9 | | 11 |

**Figure 6.** A decompression example. Each code is translated by recursive replacement of the code with a prefix code and extension character from the string table (Table 3). For example, code 5 is replaced by code 2 and a, and then code 2 is replaced by b.

From: http://www.dspguide.com/graphics/T_27_3.gif

**FIGURE 27-6**

Example of code table compression. This is the basis of the popular LZW compression method. Encoding occurs by identifying sequences of bytes in the original file that exist in the code table. The 12 bit code representing the sequence is placed in the compressed file instead of the sequence. The first 256 entries in the table correspond to the single byte values, 0 to 255, while the remaining entries correspond to *sequences* of bytes. The LZW algorithm is an efficient way of generating the code table based on the particular data being compressed. (The code table in this figure is a simplified example, not one actually generated by the LZW algorithm).

Example Code Table

| code number | translation |
|---|---|
| 0000 | 0 |
| 0001 | 1 |
| ⋮ | ⋮ |
| 0254 | 254 |
| 0255 | 255 |
| 0256 | 145  201  4 |
| 0257 | 243  245 |
| ⋮ | ⋮ |
| 4095 | xxx  xxx  xxx |

*identical code* (0000–0255)
*unique code* (0256–4095)

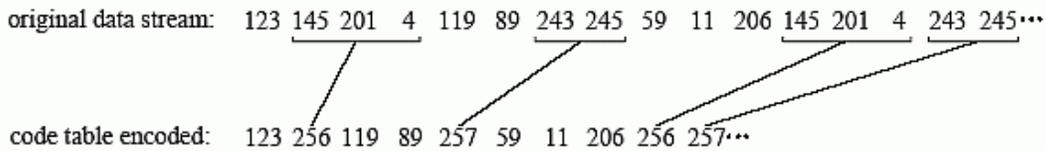

original data stream:    123  145  201   4   119  89  243  245  59   11  206  145  201   4   243  245···

code table encoded:      123  256  119  89  257  59   11  206  256  257···

| | CHAR | STRING + CHAR | In Table? | Output | Add to Table | New STRING | Comments |
|---|---|---|---|---|---|---|---|
| 1 | t | t | | | | t | first character- no action |
| 2 | h | th | no | t | 256 = th | h | |
| 3 | e | he | no | h | 257 = he | e | |
| 4 | / | e/ | no | e | 258 = e/ | / | |
| 5 | r | /r | no | / | 259 = /r | r | |
| 6 | a | ra | no | r | 260 = ra | a | |
| 7 | i | ai | no | a | 261 = ai | i | |
| 8 | n | in | no | i | 262 = in | n | |
| 9 | / | n/ | no | n | 263 = n/ | / | |
| 10 | i | /i | no | / | 264 = /i | i | |
| 11 | n | in | yes (262) | | | in | first match found |
| 12 | / | in/ | no | 262 | 265 = in/ | / | |
| 13 | S | /S | no | / | 266 = /S | S | |
| 14 | p | Sp | no | S | 267 = Sp | p | |
| 15 | a | pa | no | p | 268 = pa | a | |
| 16 | i | ai | yes (261) | | | ai | matches *ai*, *ain* not in table yet |
| 17 | n | ain | no | 261 | 269 = ain | n | *ain* added to table |
| 18 | / | n/ | yes (263) | | | n/ | |
| 19 | f | n/f | no | 263 | 270 = n/f | f | |
| 20 | a | fa | no | f | 271 = fa | a | |
| 21 | l | al | no | a | 272 = al | l | |
| 22 | l | ll | no | l | 273 = ll | l | |
| 23 | s | ls | no | l | 274 = ls | s | |
| 24 | / | s/ | no | s | 275 = s/ | / | |
| 25 | m | /m | no | / | 276 = /m | m | |
| 26 | a | ma | no | m | 277 = ma | a | |
| 27 | i | ai | yes (261) | | | ai | matches *ai* |
| 28 | n | ain | yes (269) | | | ain | matches longer string, *ain* |
| 29 | l | ainl | no | 269 | 278 = ainl | l | |
| 30 | y | ly | no | l | 279 = ly | y | |
| 31 | / | y/ | no | y | 280 = y/ | / | |
| 32 | o | /o | no | / | 281 = /o | o | |
| 33 | n | on | no | o | 282 = on | n | |
| 34 | / | n/ | yes (263) | | | n/ | |
| 35 | t | n/t | no | 263 | 283 = n/t | t | |
| 36 | h | th | yes (256) | | | th | matches *th*, *the* not in table yet |
| 37 | e | the | no | 256 | 284 = the | e | *the* added to table |
| 38 | / | e/ | yes | | | e/ | |
| 39 | p | e/p | no | 258 | 285 = e/p | p | |
| 40 | l | pl | no | p | 286 = pl | l | |
| 41 | a | la | no | l | 287 = la | a | |
| 42 | i | ai | yes (261) | | | ai | matches *ai* |
| 43 | n | ain | yes (269) | | | ain | matches longer string *ain* |
| 44 | / | ain/ | no | 269 | 288 = ain/ | / | |
| 45 | EOF | / | | / | | | end of file, output *STRING* |

TABLE 27-3
LZW example. This shows the compression of the phrase: *the/rain/in/Spain/falls/mainly/on/the/plain/*.